\documentclass[12pt]{article}
\usepackage{latexsym,epsfig,graphicx,epstopdf,amsmath,amssymb,amscd,undertilde,multirow,chicago,psfrag,paralist,dsfont,url}
\usepackage[titletoc]{appendix}

\usepackage[american]{babel}

\newcommand{\thetavec}{{\boldsymbol{\theta}}}

\newcommand{\zvec}{{\boldsymbol{z}}}

\newcommand{\betavec}{{\boldsymbol{\beta}}}

\newcommand{\xvec}{\boldsymbol{x}}

\newcommand{\muvec}{\boldsymbol{\mu}}

\begin{document}

\title{Applied Statistics in the Era of Artificial Intelligence: \\ A Review and Vision}

\author{
Jie Min$^{1}$, Xinyi Song$^{2}$, Simin Zheng$^{2}$, Caleb B. King$^{3}$, \\ Xinwei Deng$^{2}$, and Yili Hong$^{2}$\\[1.5ex]
{\small $^1$Department of Mathematics \& Statistics, University of South Florida, Tampa, FL 33620}\\
{\small $^{2}$Department of Statistics, Virginia Tech, Blacksburg, VA 24061}\\
{\small $^{3}$JMP Division, SAS, Cary, NC 27513}
}

\date{}

\maketitle

\begin{abstract}

The advent of artificial intelligence (AI) technologies has significantly changed many domains, including applied statistics. This review and vision paper explores the evolving role of applied statistics in the AI era, drawing from our experiences in engineering statistics. We begin by outlining the fundamental concepts and historical developments in applied statistics and tracing the rise of AI technologies. Subsequently, we review traditional areas of applied statistics, using examples from engineering statistics to illustrate key points. We then explore emerging areas in applied statistics, driven by recent technological advancements, highlighting examples from our recent projects. The paper discusses the symbiotic relationship between AI and applied statistics, focusing on how statistical principles can be employed to study the properties of AI models and enhance AI systems. We also examine how AI can advance applied statistics in terms of modeling and analysis. In conclusion, we reflect on the future role of statisticians. Our paper aims to shed light on the transformative impact of AI on applied statistics and inspire further exploration in this dynamic field.


\end{abstract}

\newpage

\section{Introduction}\label{sec:introudction}
\subsection{Basic Workflow of Applied Statistics}

\begin{figure}[h]
\centering
\includegraphics[width=.95\textwidth]{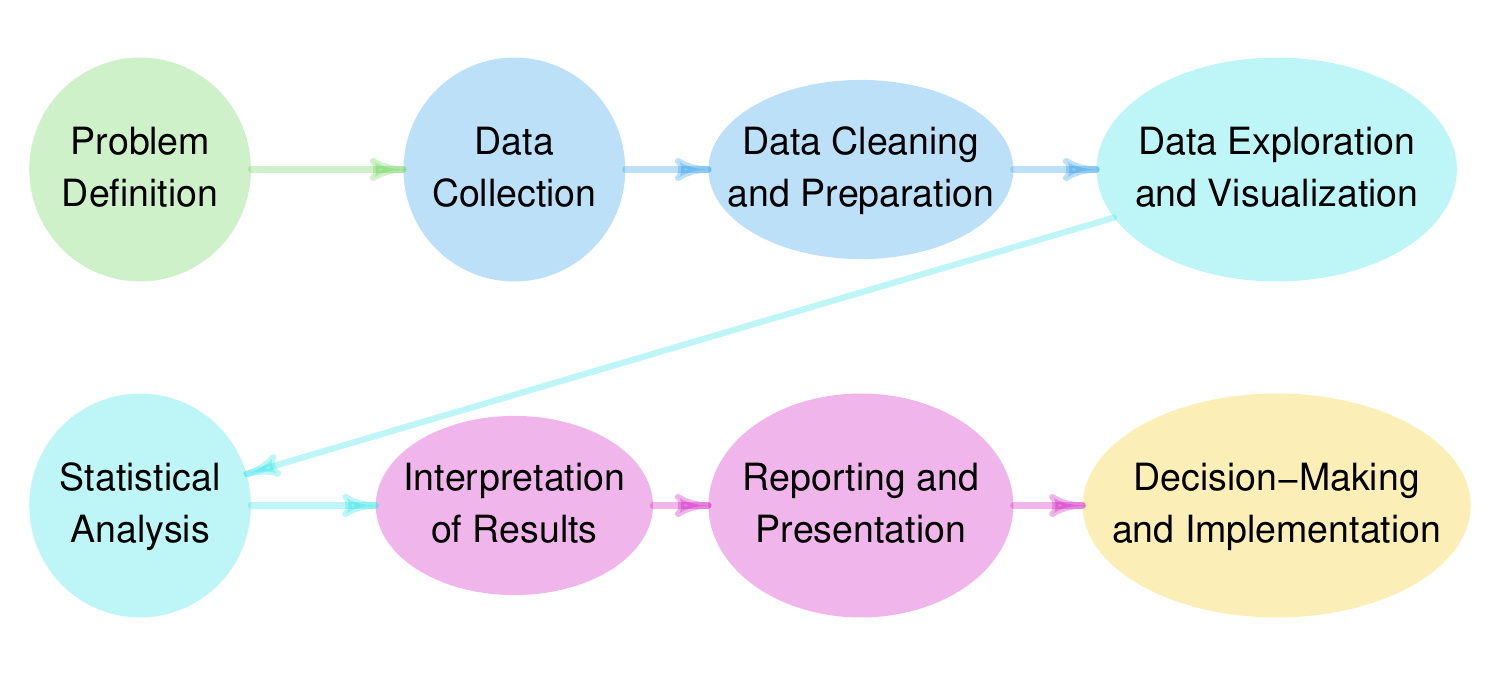}
\caption{Flowchart for the main steps in applied statistics.}\label{fig:apllied.stat.flowchart}
\end{figure}

Applied statistics involves the use of statistical methods and theories to analyze real-world data, drawing insights and making informed decisions across various fields such as business, healthcare, engineering, social sciences, and more. While the process can vary across applications, the basic workflow of applied statistics consists of a systematic process of gathering, analyzing, interpreting, and presenting data. In this paper, we outline eight main steps to illustrate the workflow of applied statistics, as shown in Figure~\ref{fig:apllied.stat.flowchart}.

This is by no means a rigid, standardized process that all applications must follow; rather, most applications will involve these steps to some extent. In the following sections, we will describe each of the eight steps depicted in Figure~\ref{fig:apllied.stat.flowchart}.

\subsubsection{Problem Definition}

The first step before any analysis can begin is to accurately and succinctly define the problem being studied. This includes an understanding of the history and context in which the problem arises. Questions considered at this stage include some of the following:

\begin{inparaitem}
\item \emph{People:} Who are the key stakeholders? Who needs to be on the project team? Are there any circumstances that might affect the analysis process and/or outcome? Who is responsible for implementing and maintaining any proposed solutions from the analysis?

\item \emph{History:} Have there been any previous attempts to address the problem? If so, are they documented, and why is a new analysis being conducted? What will be done differently this time?

\item \emph{Technology \& Resources:} What resources are available to the team? What technical expertise is needed and available? Are there any constraints to be aware of?

\item \emph{Problem Scope:} What is the extent of the problem? Is a local solution sufficient to start? What are potential downstream impacts?
\end{inparaitem}

Answering these questions helps the team or individual conducting the analysis to properly identify the context surrounding the problem, which will then guide the development of their plan. A key part of this plan is the definition of clear metrics which allow the team or individual to assess their progress and ultimately determine the success of their plan.

\subsubsection{Data Collection, Cleaning, and Preparation}

Once the problem has been appropriately defined, the next stage is to collect the \emph{necessary} data. We emphasize ``necessary" as there is often a temptation by analysts and researchers to simply gather any and all data that may be even tangentially related and then filter out unnecessary data as the project proceeds, especially if resources are plentiful. However, as an article in the Harvard Business Review succinctly puts it, ``You Don't Need Big Data - You Need the Right Data" (\citeNP{Wessel2016}). Data collection is a time and resource intensive process. If poorly planned, it can result at best in an unnecessarily longer data cleaning period and at worse misinformed analyses and models leading to spurious results. The experimental design principles, such as randomization, blocking, and replication, are very important for data collection (\citeNP{Box1978statistics}). Methods such as factor screening and designed data collection (\citeNP{AndersonCookLu2023}) are essential for ensuring high data quality and optimal use of resources.

After the data have been collected, there still remains the task of cleaning and preparing the data for analysis. This typically involves detecting and addressing outliers, making any necessary transformations to the data, and data formatting. A carefully constructed data collection procedure will reduce the time needed for data cleaning and preparation, especially if it can be done in parallel with data collection.  A quote from Dennis Lin of Purdue University summarizes this step well: ``It's not about getting the data right, it's about getting the right data".

\subsubsection{Data Exploration and Visualization}

The next stage in the analysis workflow involves data exploration and visualization. One of the key uses of data visualization is to assess the underlying assumptions typically required for many statistical methods. Anscombe's quartet (\citeNP{Anscombe1973}) and the Datasaurus dozen (\citeNP{MatejkaFitzmaurice2017}) are simple, yet powerful illustrations of this point, each providing a series of vastly different data patterns that yet all yield the same statistical results. Data exploration is also encouraged as it can inform new paths for analysis. Some visualizations of the data may lead to new insights and new opportunities for investigation. The right visualization can sometimes provide more information than any statistical test or model, as is often illustrated in the work of Tufte (see \citeNP{Tufte1997} and \citeNP{Tufte2020}).

\subsubsection{Statistical Analysis \& Interpretation of Results}

We now come to the stage most often associated with applied statistics. It is here that the patterns and trends observed in the previous stage are either confirmed or supplanted with statistical rigor. Here, the team or individual conducting the analysis must avoid the temptation to place too much emphasis on seeking out trends and new relationships. The recent reproducibility crisis in scientific research is a cautionary tale of what can happen when disregarding rigor in favor of novel findings to drive publications. It is the job of the analyst and their team to understand the limitations of the tools in their statistics and data science toolsets and embrace the uncertainty that comes with working with data. Sometimes the results are inconclusive and it is perfectly acceptable to acknowledge the need for more data or additional context. A well-formed data collection scheme may even be able to take this into account. The analysis workflow is not a linear flow, but rather a turbulent one, twisting back on itself repeatedly as new observations and insights are revealed.

\subsubsection{Reporting and Presentation}

Once the analyses are concluded, the next stage is to compile a thorough, yet concise report and present on the findings. This is especially important if the intended recipients comprise a variety of backgrounds. A good report will be sparing on the mathematical details of the analysis, yet be transparent on any assumptions made in the process. Visualizations are of vital importance to communicating the results, showing again the importance of data exploration and visualization.

\subsubsection{Decision-Making and Implementation}

The final stage in the applied statistics workflow is the implementation of the team or individual's findings. Even if they are not ultimately responsible for this implementation, it is important that this end goal always be in mind during the previous stages. A motivating example is the oft-cited Netflix competition, in which the winning team's model was never implemented due to time and cost constraints. While the team certainly benefited financially (the prize was \$1,000,000), they ultimately failed in their task of providing Netflix with the solution for which the competition was set up in the first place. All of the best analytics, algorithms, and visualizations are ultimately meaningless if they do not result in permanent improvement to an organization or process.

\subsection{History of Thoughts}\label{three areas}

Within the past two decades, the scope and complexity of data have dramatically increased with the development of modern technology. Big data has become a popular topic, which brings many changes and opportunities to applied statistics, along with a few challenges. Before discussing further, we briefly introduce the historical development of applied statistics, and the relationship between applied statistics and two closely related fields: data analytics and data science.

Basic statistics, such as frequency counts and other summary measures, have been used for millennia. Much of what we may recognize today as applied statistics was derived from the work of notable individuals in the 1600s and 1700s, such as John Graunt and his compiling of life data into mortality tables (\citeNP{Wilcox1938}) and the development of the least squares technique for use with astronomical studies (\citeNP{Stigler1990}). Applied statistics generally focuses on the rigorous application of statistical methodologies to solve specific problems, often with a strong emphasis on theoretical foundations. As the name implies, it is in essence the application of what is developed in the discipline of mathematical statistics, while also influencing such development.

Over time, there have been multiple recommendations on how to extend the scope and impact of statistics. \citeN{HoerlSnee2010a} highlighted the importance of statistical thinking and statistical engineering, which studies how to utilize and develop statistics frameworks to  benefit other scientific disciplines. \citeN{HoerlSnee2010} raised several issues in the development of statistics in the context of quality improvement, and suggested that practical utilization of statistics should receive more attention. \citeN{Kenett2015} introduced a life cycle view of statistics, including defining the problems and goals, data collection and analysis, implementation of findings, communication, and impact assessment, most of which are depicted in Figure~\ref{fig:apllied.stat.flowchart}. They suggested that statistics is able to have larger impact toward research and industry using the life cycle view together with quality evaluation of information generated by statistical analysis.

Data analytics is a discipline that centers on practical data examination to generate insights for informed decision-making, often within a business context. It began as a special topic of applied statistics starting around the 1950s with a strong focus on business applications (\citeNP{AyoVaughan2023}). Since then, it has developed a greater emphasis on data management and storage; an area that can often be overlooked in applied statistics. And although it might seem the youngest of the three disciplines, data science is actually nearly as old as data analytics. Data science can be traced back to John Tukey's work in \citeN{Tukey1962} in what he foresaw as a better description of the analytical process that would arise in the near future with more automated procedures. More recently, \citeN{Donoho2017} claimed the core of data science lies in learning from the data, and introduced the full scope of data science, covering both statistics and machine learning. \citeN{SteinbergAronovich2020} further explored the usage of data science in business and industry, and argued it is essential that data science contributes to improving the performance of companies in the industry, rather than focusing only on problems driven by intellectual challenge.

Data science can be seen as a comprehensive approach to data analysis, including statistical, computational, and machine learning methods to handle large datasets and solve complex problems across various domains. As compared to applied statistics, there is a greater emphasis on taking advantage of computation methods in the data analysis process. And with the rise of AI and machine learning, this discipline has truly come into its own. This is evidenced by its use applications such as natural language processing (NLP), statistical data mining (\shortciteNP{ribeiro2017importance}), advanced data visualization, and parallel computing.

In presenting these disciplines, it is clear that each has a distinct role but overlaps significantly, especially in their goal to extract valuable insights from data. In particular, each tends to emphasize some part of the data analysis process. Applied statistics values the underlying theory that justifies many of the techniques we may take for granted. Data analytics sees the value in proper data management and access. Finally, data science has clearly demonstrated the advantages offered by computational technology. Rather than view them as competing methodologies, a better approach is to view these fields as complementary, each ensuring certain portions of the data analysis process is not overlooked.

\subsection{The Emergence of Artificial Intelligence}

Alongside the rise of big data and data science, there has also been tremendous growth in research and applications of artificial intelligence (AI) in the past decade (\shortciteNP{srivastava2024exploring}). Creating intelligent artificial beings has long been a dream of humankind, tracing back to ancient history. Modern AI has its origins in the 20th century, when \citeN{turing1936computable} introduced the concept of a universal machine capable of computing anything that is computable, laying the groundwork for digital computers. Turing later developed the test named after him in order to evaluate a machine's ability to exhibit behavior indistinguishable from that of a human (\citeNP{turing1950}). The name ``artificial intelligence'' was officially coined at a Dartmouth conference in 1956, marking the official birth of AI as a field of research (\shortciteNP{McCarthy_Minsky_Rochester_Shannon_2006}).

Over the following decades, the development of AI experienced periods of rapid progress (booms) and stagnation (winters), yet through it all significant advancements were made. Among these was the development of neural networks, inspired by the functioning of neurons in the human brain. \citeN{mcculloch1943logical} initially laid the groundwork by deriving the concept of artificial neurons. The idea of the perceptron, introduced by \citeN{rosenblatt1958perceptron}, demonstrated how neural networks could be used for classification and learning from data. Later, perceptron models were expanded to multilayer networks, with \shortciteN{rumelhart1986learning} introducing the concept of back-propagation for training multilayer networks, thus enabling the creation of more complex and effective neural networks.

Moving into the 2010s, we witnessed a revolution in big data and computing hardware, particularly with the advent of more advanced graphics processing units (GPUs). This period marked the rise of deep learning models, which demonstrated significant breakthroughs and applications across various fields. For example, AlexNet showcased the transformative power of deep learning in image recognition (\shortciteNP{krizhevsky2012imagenet}). \shortciteN{goodfellow2014generative} introduced generative adversarial networks, enabling the generation of synthetic data and images.  The bidirectional encoder representations from transformers (BERT) model, developed by \shortciteN{devlin2019bert}, advanced NLP by allowing machines to better understand context in language. Additionally, the generative pre-trained transformer (GPT) model (\shortciteNP{radford2018improving}) demonstrated remarkable proficiency in generating human-like text, answering questions, and performing a wide range of language tasks.

As of today, the latest ChatGPT models are capable of engaging in coherent and contextually relevant dialogues while assisting with a wide range of tasks across various domains. The ongoing development of AI presents new opportunities for statisticians, including access to more powerful tools for statistical analysis and research avenues in explainable AI, uncertainty quantification, and AI safety. These opportunities will be explored further in this paper.

The rest of the paper is organized as follows. Sections~\ref{sec:tradition.areas.AS} and~\ref{sec:emerging.areas.AS} introduce traditional and emerging areas of applied statistics with examples from engineering statistics. Section~\ref{sec:AS.for.AI} discusses how applied statistics techniques can be used for AI development. Section~\ref{sec:AI.for.AS} discusses how AI techniques can be used for statistical analysis. Section~\ref{sec:concluding.remarks} discusses the future of statisticians and concludes the paper with some remarks.

\section{Traditional Areas of Applied Statistics}\label{sec:tradition.areas.AS}
\subsection{Traditional Areas}

Applied statistics has proven invaluable across diverse fields, including medicine, biology, engineering, and business. In the medical and biological domains, biostatistics plays a key role in analyzing biological and health-related data, supporting tasks such as understanding and treating diseases and developing new drugs. Traditional methods used in biostatistics include linear and generalized linear models, analysis of variance, mixed-effects models, and hypothesis testing (\citeNP{pagano2022principles}). In clinical trials, survival models are used to analyze time-to-event data (\citeNP{collett2023modelling}), while longitudinal data analysis is employed for repeated measures data. In bioinformatics, statistical methods aid in variable selection (\citeNP{zhang2024robust}) and hypothesis testing.

In engineering statistics, statistical methods are used and developed to ensure and improve quality in manufacturing processes as well as assess reliability of products. In the latter case, statistical models are developed to analyze failure events and predict the future failure time of products. In business and industry, applied statistics are useful for decision-making and forecasting. For example, time series analysis can be used to predict stock price (\shortciteNP{kumar2022systematic}), logistic regression can be used in credit risk analysis, and A/B testing is helpful in evaluating the effectiveness of a new product.

In the era of AI, these traditional areas of applied statistics have been influenced and inspired by new AI and machine learning models. In the following sections, we provide two examples from our research to show how applied statistics are used to solve problems from engineering areas. We present these examples according to the workflow shown in Figure~\ref{fig:apllied.stat.flowchart}.

\subsection{Reliability Analysis of GPU in Supercomputers}
GPUs are widely used in supercomputers to train AI models, and the reliability of GPUs inside supercomputers is an important research topic. To analyze the reliability of GPUs, a spatially correlated competing risks model is proposed in \shortciteN{Min2023-GPU}.

\begin{figure}
	\centering
	\begin{tabular}{ccc}
		\includegraphics[width=.3\textwidth]{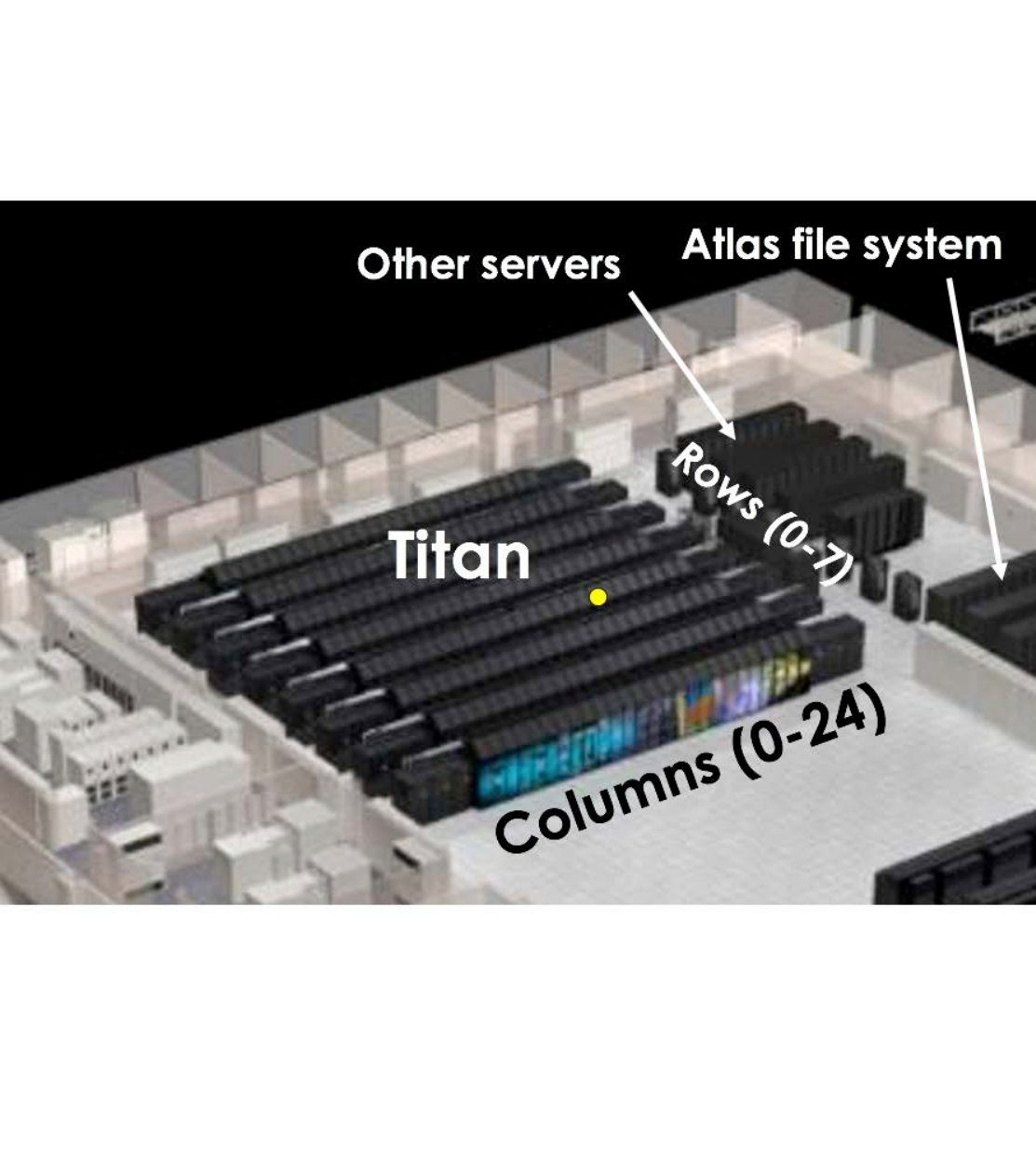} &
		\includegraphics[width=.3\textwidth]{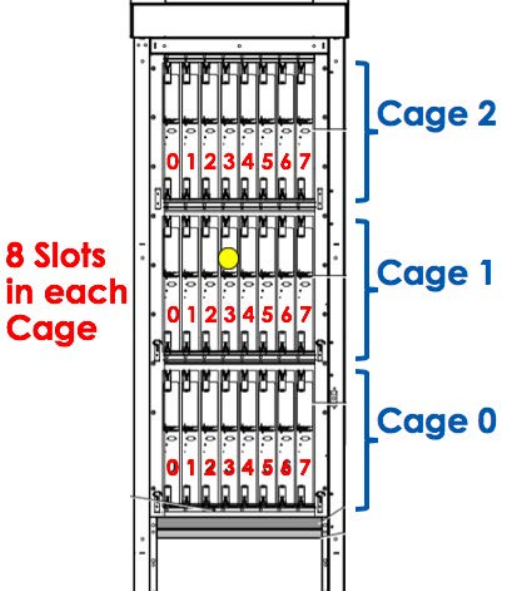}&
		\includegraphics[height=.3\textwidth]{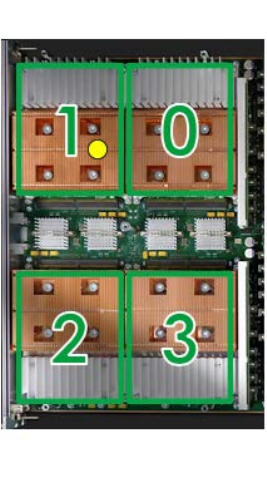} \\
		(a) Cabinet Array &
		(b) Cabinet Structure &
		(c) Slot Structure \\
	\end{tabular}
	\caption{Illustration of the layout of Titan supercomputer. \emph{Figure reproduced with permission from Ostrouchov et al. (2020)}.}\label{fig:tts}
\end{figure}

\begin{figure}
	\begin{center}
			\includegraphics[width=0.9\textwidth]{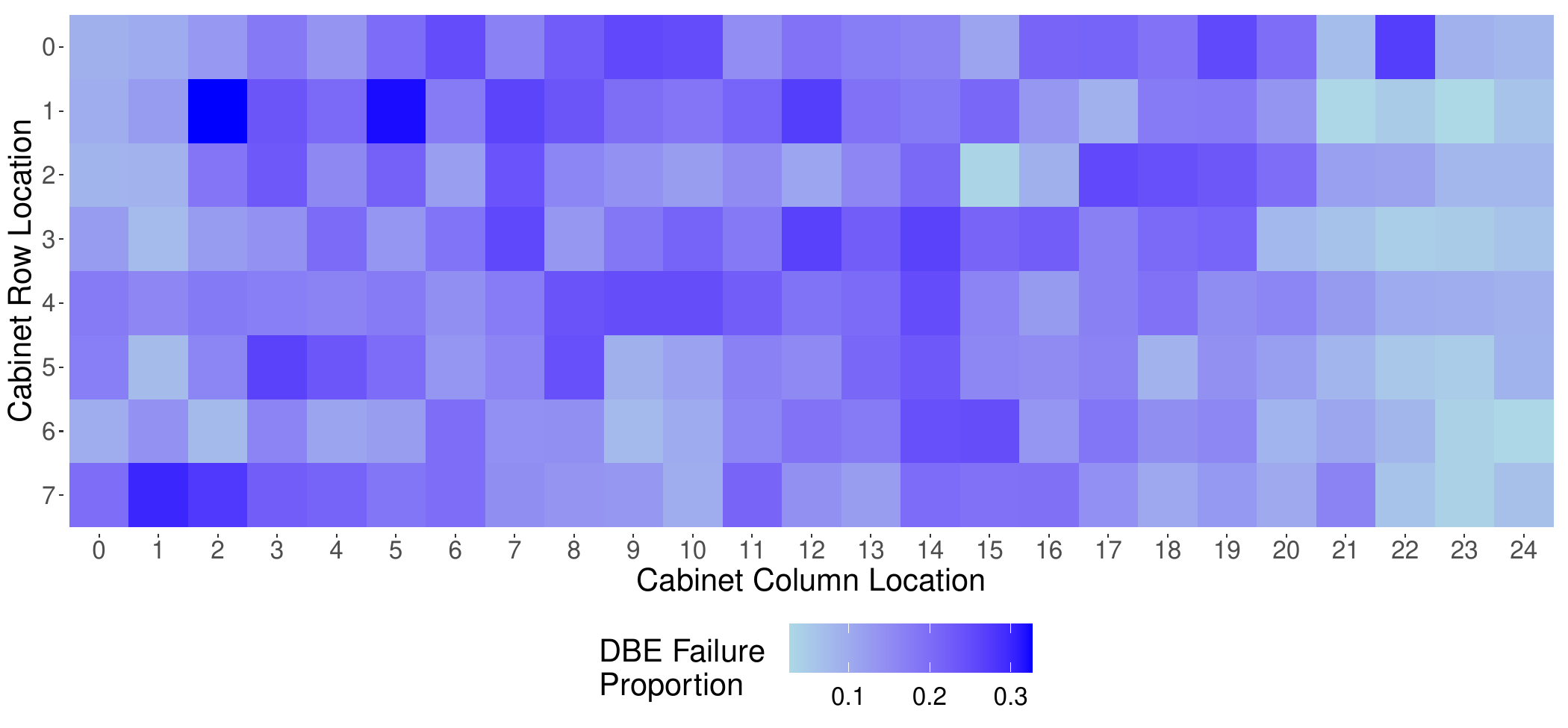}
		\caption{Heatmap for the DBE failure proportion over the $8\times 25$ spatial domain.}\label{fig:dbefr}
	\end{center}
\end{figure}

The GPU failure time data were collected from the Cray XK7 Titan supercomputer, and were made available to the public by \shortciteN{Ostrouchovetal2020}. Figure~\ref{fig:tts} shows the architecture of the Titan supercomputer. The supercomputer has 25 columns and 8 rows of cabinets, with 3 cages inside each cabinet, 8 slots inside each cage, and 4 nodes inside each slot. One GPU is inserted on each node. In Titan, there are two major errors (i.e., failure events) that can lead to GPU replacement: the double bit (DBE) error (i.e., the correction of a single-bit flip, or the detection of a double-bit flip) and off-the-bus (OTB) error (i.e.,  the loss of host CPU connection to the GPU). Figure~\ref{fig:dbefr} shows the proportion of the DBE failures on 200 cabinet locations, which indicates potential spatial correlation among the failures.

In the statistical analysis, cage, slot, and node factors in the supercomputer are used as covariates, and spatially correlated random effects are used to capture the information from the cabinet locations. The random effects of DBE and OTB errors on $200$ locations are modeled by a normal distribution with covariance matrix $\Sigma_f \otimes \Omega$, where $\Omega$ describes the spatial correlation among the cabinet locations. A special distance function is proposed based on both the cabinet physical locations and the cabinet logical connections. Furthermore, correlated competing risks of the DBE and OTB errors is modeled using the $2\times2$ matrix $ \Sigma_f$.

\shortciteN{Min2023-GPU} concludes that both the cabinet locations, and the GPU positions inside the cabinets have influence on the GPU failure times. The analysis results also suggest a high correlation between OTB and DBE failure times. The results shed lights on the failure mechanisms of GPU and can be useful in designing future supercomputer systems.

	\begin{figure}
	\begin{center}
		\begin{tabular}{cc}
			\includegraphics[width=0.45\textwidth]{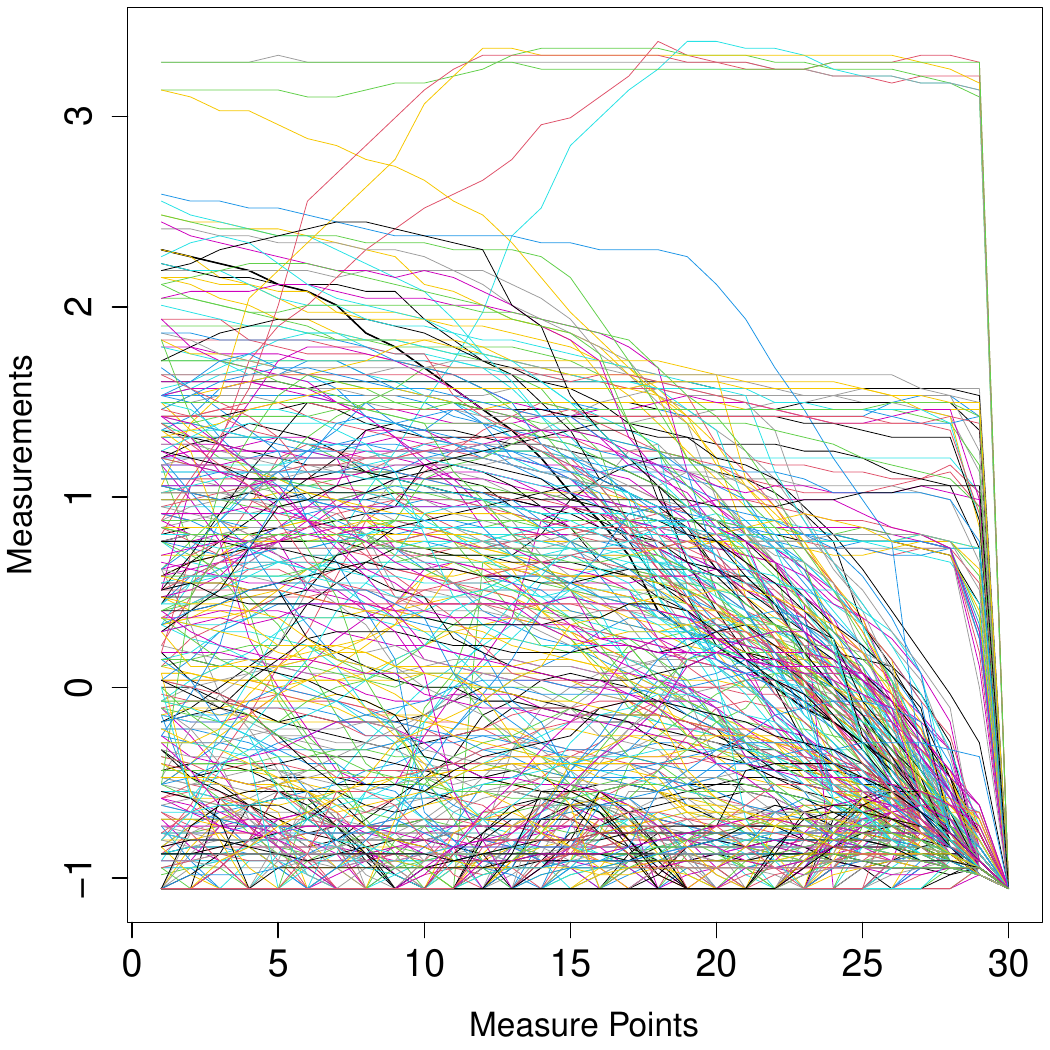}&
			\includegraphics[width=0.45\textwidth]{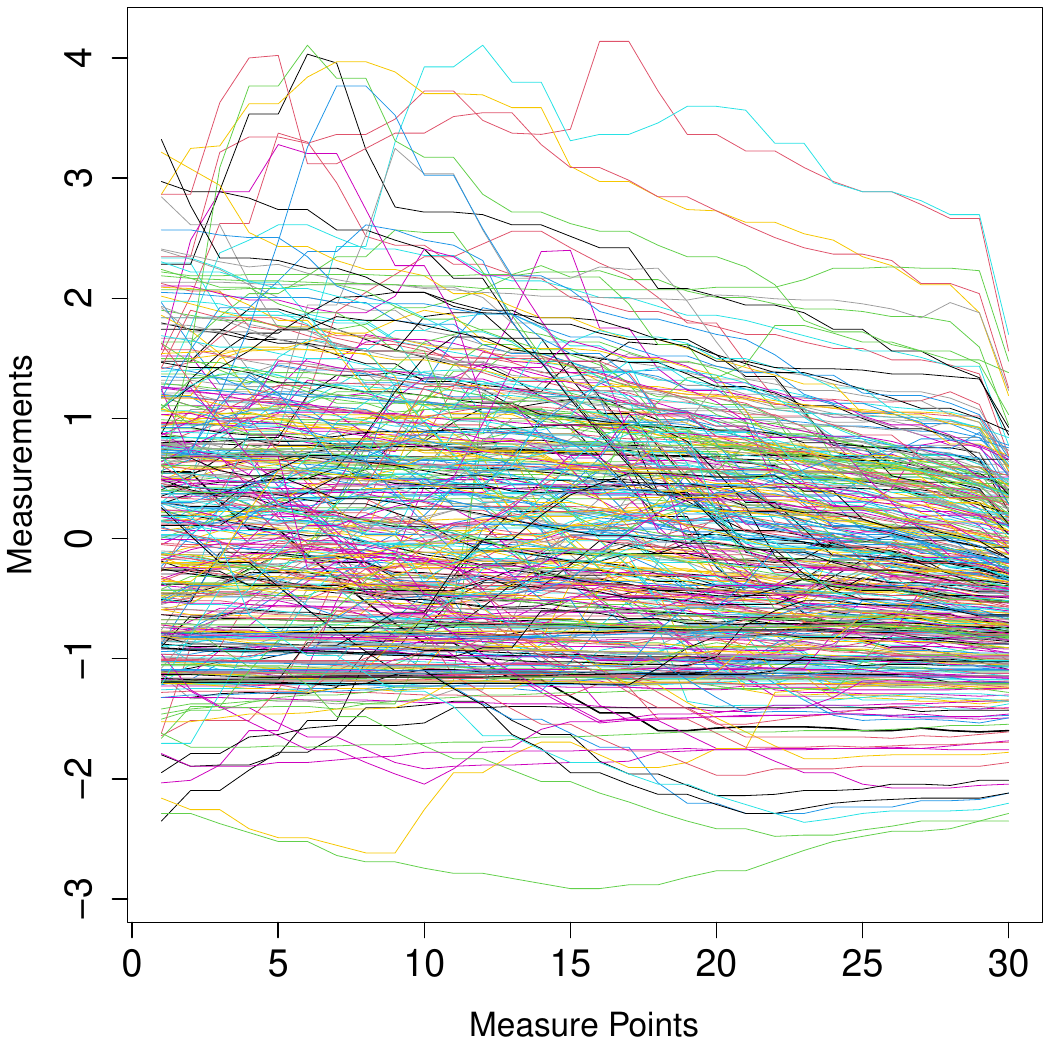}\\
			(a) Sensor 1 & (b) Sensor 2
		\end{tabular}
		\caption{Examples of data from two sensors, with each line shows a trajectory from the senor. \emph{Figure reproduced with permission from INFORMS}.}\label{fig:engineerSensorExample}
	\end{center}
\end{figure}

\subsection{Sensor Data Clustering with Variable Selection}

Modern engineering systems often have multiple sensors installed to monitor the status of the systems. Extracting information from sensor data has become a popular research topic within the context of big data and both statistics and machine learning algorithms contribute in this area. For example, \shortciteN{Wangetal2022RESS} used a machine learning approach to construct a degradation index from sensor data while \shortciteN{Jinetal2024} took a more statistical approach in developing a clustering method with variable selection for sensor data.

Here we give a brief introduction to \shortciteN{Jinetal2024} to showcase the use of statistical methods for sensor data. Figure~\ref{fig:engineerSensorExample} shows the data from two sensors from an engineering system. Because the sensor data can be viewed as functional curves, before clustering, a functional principal analysis is first performed for data reduction. The first several functional principal components are chosen to be used in clustering for each sensor and observation. Model-based clustering with variable selection is then performed based on the principal component coefficients.

\begin{figure}
	
	\begin{center}
		\includegraphics[width=0.55\textwidth]{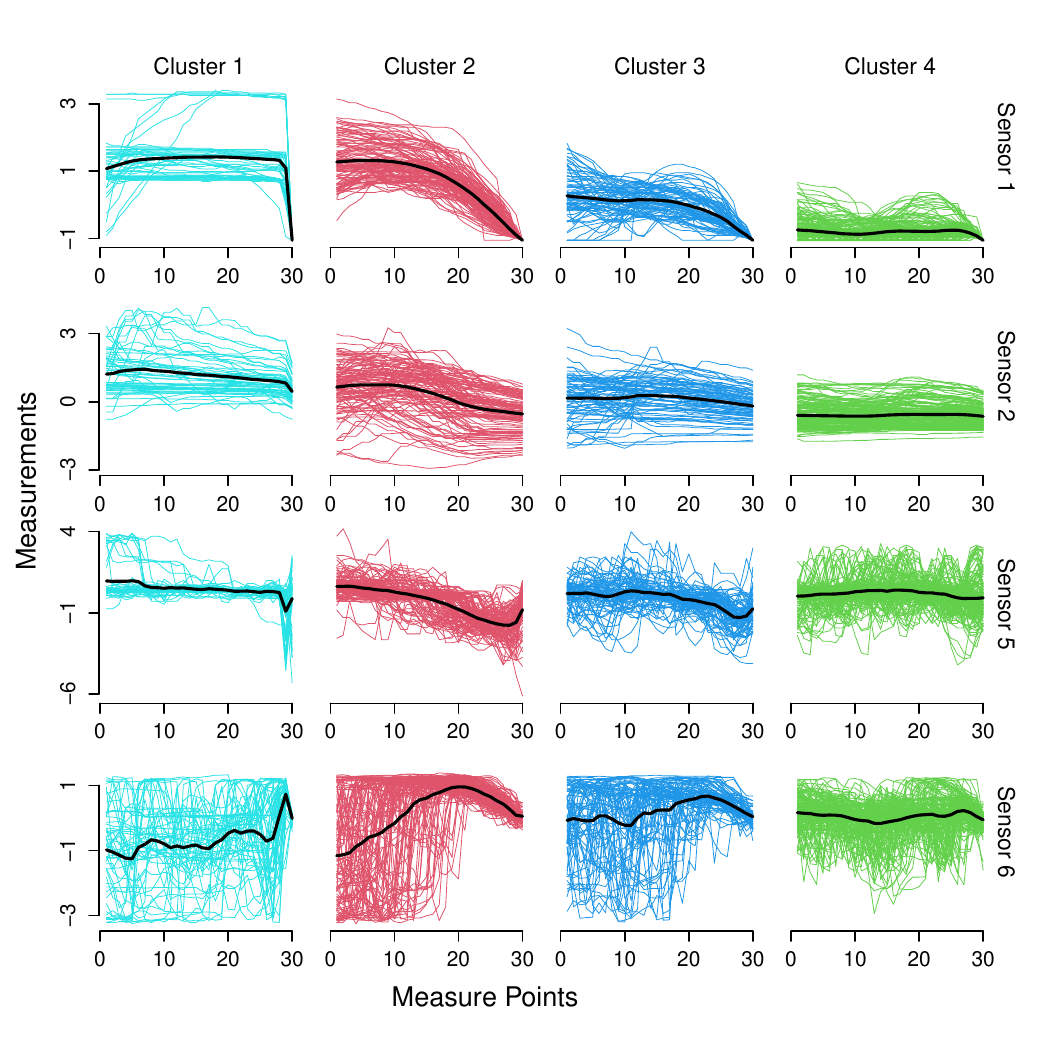}
		\caption{Illustration of clustering results based on data from the four sensors stayed in the model. \emph{Figure reproduced with permission from INFORMS}.}
		\label{fig:selectedsensor}
	\end{center}	
\end{figure}

\shortciteN{Jinetal2024} conducted variable selection by using a penalty scheme. Figure~\ref{fig:selectedsensor} shows the clustering results for four non-removed sensors based on the engineering sensor data. The clustering result shows different behaviors of the sensor data in different clusters. The proposed method can be used to recognize important sensors that are related to the event of interest, and to improve the performance of machine learning models.

\section{Emerging Areas of Applied Statistics}\label{sec:emerging.areas.AS}
\subsection{A Brief Introduction}

In addition to the traditional areas of applied statistics discussed in Section~\ref{sec:tradition.areas.AS}, the advancement of technology has given rise to some emerging areas in applied statistics. Statistical methods have been developed to evaluate the performance of high-performance computing (HPC), assess reliability in renewable energy technology, analyze social networks, study the properties of new materials, and improve quality in additive manufacturing (i.e., 3D printing). The following sections present several examples from these emerging areas of engineering statistics.

\subsection{Applications in HPC Performance Study}

Recently, there has been increasing attention on the performance variability of HPC systems (e.g., \shortciteNP{Wangetal2023JQT} and \shortciteNP{Xuetal2024-JRSSC}). Research on HPC performance and its variability provides tremendous opportunities for applied statisticians. The central questions are (1) how to measure the performance and performance variability, (2) what are the main factors that affect performance variability, and (3) what is the performance prediction under a certain configuration of a system. For example, \shortciteN{Xuetal2024-JRSSC} proposed a modified Gaussian process (GP) that can be used to predict the performance distribution of HPC systems with experimental data from HPC input/output tasks used to illustrate the proposed methods.

\begin{figure}
    \centering
    \includegraphics[width = 0.65\linewidth]{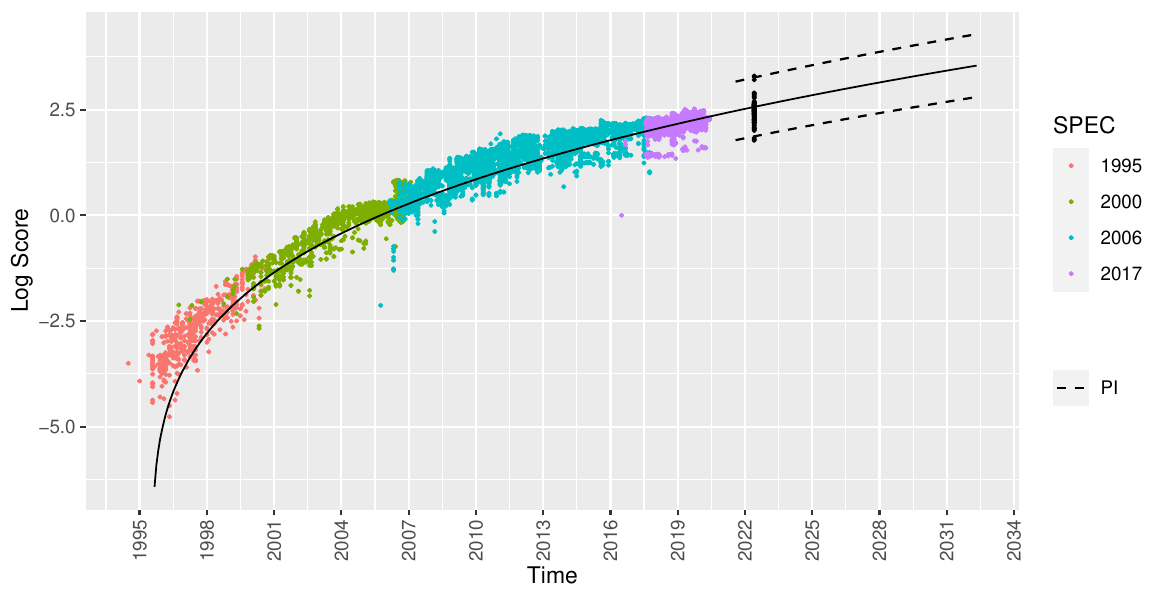}
    \caption{Visualization of the estimated mean trend, point predictions, and prediction interval (PI) for the base integer speed. \emph{Figure reproduced with permission from IEEE}. }
    \label{fig:overall.ind.trend}
\end{figure}

Predicting future computer system performance is also of interest and something that statistics is ideally suited. To this end, \shortciteN{Wangetal2024_HPC} conducted historical and statistical analyses based on over four decades of HPC performance data from the iLORE database. Specifically, they performed both isolated and collective sensitivity analyses, and then predicted future performance based on historical data. The fitted mean trend, point predictions, and the prediction intervals of the overall score for the next 100 months are shown in Figure~\ref{fig:overall.ind.trend}. From this, we can see a clear upward trend in the log score over time, indicating that computing performance, as measured by benchmarks, has been improving consistently from 1995 to 2017, and is expected to continue improving in the future.

\subsection{Applications in Renewable Energy}

Another emerging field of engineering statistics is on renewable energy. Rechargeable lithium-ion or ``Li-ion" batteries have become a popular form of energy storage, and predicting their lifespan has gained substantial interest in statistical research. For example, \shortciteN{do2022reliability} analyzed the degradation of rechargeable Li-ion batteries through a two-step predictive modeling approach that integrates both longitudinal and functional data analysis techniques. The data used in their study originate from the NASA Prognostics Data Repository. To assess battery reliability, \shortciteN{do2022reliability} proposed a functional degradation model to evaluate battery life, with voltage discharge curves representing the degradation over time.

Figure~\ref{fig:overall.ind.trend} illustrates the predicted degradation paths for four representative batteries, using 75\% of each battery's training dataset. The figure demonstrates that the proposed functional degradation models (i.e., FDM-LME and FDM-FLMM) effectively capture the testing trajectories of the degradation paths, showing strong predictive performance. This highlights the potential of functional degradation models for accurate and reliable degradation predictions in Li-ion batteries. Furthermore, the model provides complete voltage discharge curves, allowing practitioners the flexibility to define customized degradation metrics based on these curves.

\begin{figure}
	\centering
	\includegraphics[width=0.65\linewidth]{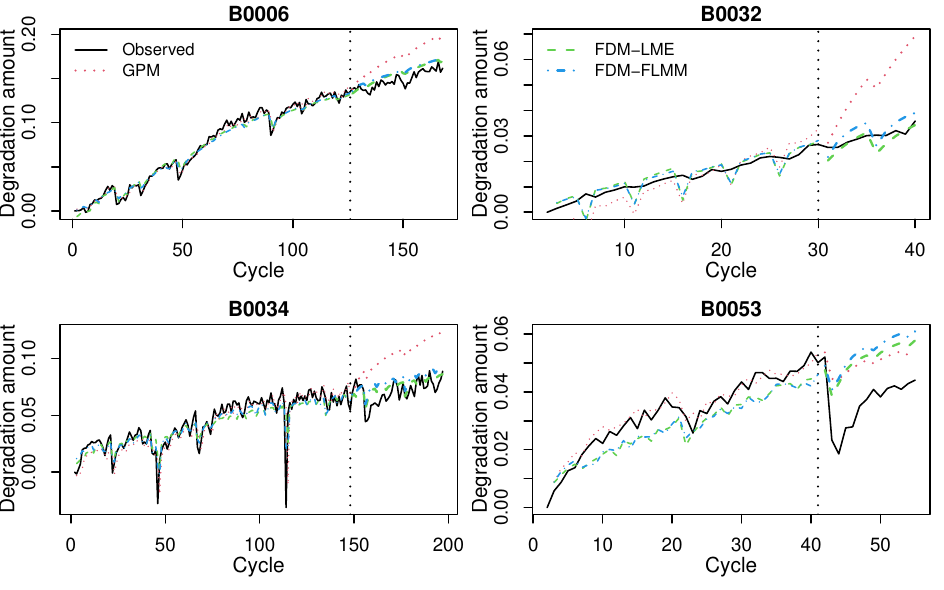}
	\caption{Visualization of model fitting and predictions based a traditional model (GPM, red lines), and new models (FDM-LME, green lines; FDM-FLMM, blue lines). \emph{Figure reproduced from Cho et al. (2024) with permission from Institute of Mathematical Statistics}.}
	\label{fig:degrad_preds75}
\end{figure}

\subsection{Other Applications}

An application area that is receiving significant attention from applied statistics is social network analysis. Social media significantly advances technology by enabling rapid information exchange, fostering global collaboration, and serving as a platform for new technologies to reach a broad audience. Accompanying this is an emerging area in applied statistics focused on the development of statistical tools in social network analysis. Statistics can play a fundamental role by providing the tools and methodological foundations to understand the structure, dynamics, and functions of networks (\shortciteNP{tabassum2018social}).

Another application area is in materials science. Polymeric materials and composites are widely used in fields such as photovoltaic (PV) products. Research on material properties and service life prediction presents many interesting statistical challenges. For example, \shortciteN{duan2017photodegradation} conducted statistical modeling and analysis of accelerated test data for photodegradation and predicted photodegradation for specimens exposed outdoors. \shortciteN{king2018reliability} introduced accelerated testing and analysis techniques for polymer materials, enabling more feasible product lifetime assessments.

Similar challenges arise in additive manufacturing, where a common goal is to refine the manufacturing process and ensure product quality and reliability. For instance, \shortciteN{wang2020statistics} presented a GP approach to characterize the dimensional quality of free-form surfaces, such as those produced through additive manufacturing processes. \shortciteN{chen2021fatigue} proposed a probabilistic physics-guided neural network to analyze the effects of various parameters on the probabilistic fatigue properties of additive-manufactured products. In this and the other areas discussed here, it is clear that applied statistics is and continues to be a valuable tool.

\section{Applied Statistics for AI}\label{sec:AS.for.AI}

We have seen the close relationship between applied statistics and modern technologies in Sections~\ref{sec:tradition.areas.AS} and~\ref{sec:emerging.areas.AS}. In Sections~\ref{sec:AS.for.AI} and \ref{sec:AI.for.AS}, we further explore how each can support and enhance the other. As noted by \citeN{Redman2024AI}, statistics and AI are highly complementary.  We begin with discussions on how statistics can be valuable for AI. As in previous sections, we provide a literature review and examples from our recent research.

\subsection{Properties of AI Models}
One key benefit that statistics can provide to AI is through study of the properties of AI models. Properties of neural network/deep-learning models have already been studied from a statistical perspective. \citeN{cheng1994neural} reviewed the importance of considering statistics in neural network research, and explained the association between neural networks and statistical approaches in regression, classification, time series analysis, etc.
\citeN{imaizumi2019deep} showed that in estimating piecewise smooth functions, the convergence rate of neural network estimators corresponds to the minimax optimal rate. \citeN{hayakawa2020minimax} indicated that deep neural network estimators have better convergence rate than linear estimators in estimating a sparse and non-convex function class. \citeN{bauer2019deep} and \citeN{schmidt2020nonparametric} showed fully connected neural networks with sigmoid and ReLU activation functions and sparsity constraints can overcome the curse of dimensionality. \citeN{kohler2021rate} further proved that the convergence rate of least square estimators of certain fully connected neural networks  without the sparsity assumptions is not related to the input dimension.

Properties of neural networks in classification have also been studied. \citeN{barron1989statistical} showed the probability of classification error using certain neural networks converges to the Bayes optimal probability of error. \citeN{kim2021fast} derived the fast convergence rate of deep neural networks with ReLU activation functions for certain kinds of true models. In addition to this research, statistics can also be useful to study other properties of AI model such as in uncertainty quantification, explainability and reliability. We will discuss those topics in the following sections.

\subsection{Uncertainty Quantification}

Understanding and quantifying uncertainty in predictions generated by AI or machine learning models is crucial, and statistics is particularly suited to this task. The Bayesian approach, particularly variational inference, is widely applied in constructing uncertainty quantification (UQ) for machine learning models. For example, \shortciteN{kwon2020uncertainty} proposed a natural method for quantifying prediction uncertainty in classification tasks using a Bayesian neural network by decomposing predictive uncertainty into two types: aleatoric (i.e., intrinsic randomness) and epistemic (i.e., systematic uncertainty). \shortciteN{olivier2021bayesian} investigated variational inference-based algorithms in training neural network, allowing users to account for both aleatoric and epistemic uncertainties.

Beyond Bayesian methods, statistical algorithms have also been developed to provide UQ in machine learning models. For instance, \shortciteN{huang2024efficient} introduced a systematic framework for assessing epistemic uncertainty that provides statistical coverage guarantees for over-parameterized neural networks in regression tasks. \shortciteN{wu2024posterior} proposed a novel approach that constructs a realistic predictive distribution by using popular data augmentation techniques, achieving UQ across various image classification datasets. Additionally, \shortciteN{jantre2024learning} presented a deep variational framework leveraging a deep generative model to quantify uncertainty in image reconstruction without requiring training data. As all of this research shows, UQ for AI models remains a dynamic area of research in statistics.

\subsection{Explainable AI}

While AI has demonstrated significant promise in modeling and prediction, this often comes at the expense of interpretability. Many AI algorithms are referred to as ``black boxes'' because it is extremely difficult, if not impossible, to discern how they arrive at their conclusions. This lack of transparency is especially concerning when these algorithms are used to inform policy decisions, medical diagnoses, or other critical areas with substantial real-world implications.

To address these concerns, there has been a push for explainable AI (XAI), which aims to develop models that not only make accurate predictions but also provide explanations of how those predictions are made (\shortciteNP{broniatowski2021psycho}). This transparency allows researchers to use these explanations to improve predictive accuracy through corrective actions. Such XAI models are often referred to as ``white box'' or ``glass box'' models.

The main advantage of XAI models is their transparency, which can alleviate many of the concerns surrounding traditional AI models. However, this transparency can sometimes come at the expense of predictive accuracy and computational speed. Research is ongoing to mitigate these drawbacks, steadily closing the gap between XAI and traditional AI models. For example, a recent family of neural network models has been developed to improve interpretability without sacrificing performance (\shortciteNP{liu2024kan}). Additional insights into XAI are available in \shortciteN{longo2024xai}.

\subsection{AI Assurance}

AI assurance is essential to ensure that AI models deliver reliable and trustworthy results. As AI models have grown in popularity in recent years, the need for AI assurance has gained increasing attention. \shortciteN{batarseh2021survey} provided an overview of various AI assurance methods, which can be applied either broadly or tailored to specific AI models. Statistics plays a critical role in the development of these AI assurance methods, as discussed in this section.

\subsubsection{AI Reliability Framework}
\begin{figure}
	\begin{center}
		\includegraphics[width=.5\textwidth]{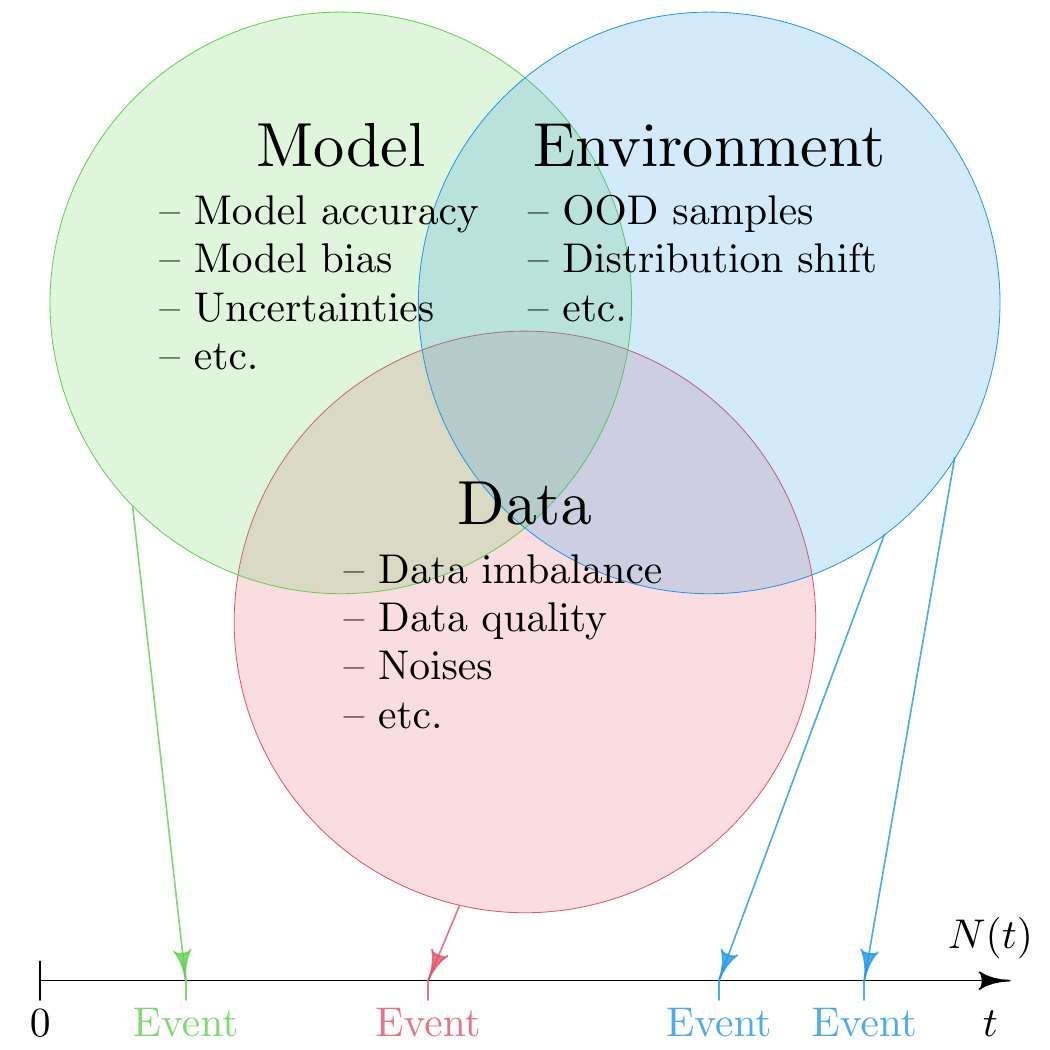}
		\caption{Possible causes of AI failures, summarized into three factors: model, environment, and data.}\label{fig:AI.rel.model.chart}
	\end{center}
\end{figure}

\shortciteN{hong2023statistical} proposed an AI reliability framework from a statistical perspective, comprising five key components: system structure, reliability metrics, failure cause, reliability assessment, and test planning. Figure~\ref{fig:AI.rel.model.chart} illustrates potential causes of AI failures based on factors from three main categories: model, environment, and data.

In reliability assessment, \shortciteN{hong2023statistical} further proposed a statistical framework to model failure events of AI systems, in which the arrival of events is modeled by counting processes. The paper also considered $k$ different types of interruptive events that can cause the failure events. The source of interruptive events includes operating environment, model, and data, as shown in Figure~\ref{fig:AI.rel.model.chart}. The arrival of interruptive events of type $j$ is then assumed to follow a counting process $N_j(t)$ with intensity function $\lambda_j[t; \xvec(t)]$, where $j = 1,2, \dots, k$, and $\xvec(t)$ contains the external failure causes such as distribution shift, data noise, and adversarial attacks. The probability that the interruptive events turn into failure events is denoted by $p_j(\zvec)$, where $\zvec$ includes internal reliability properties of the AI systems, for example, the potential of out-of-distribution detection. Then, the intensity function of $N(t)$ for failure events is
\begin{equation}
	\lambda[t; \xvec(t), \zvec] = \sum_{j = 1}^k \lambda_j[t; \xvec(t)] p_j(\zvec; \betavec_j),\label{eq:lambda1}
\end{equation}
where $\betavec_j$ contains the coefficients corresponding to $\zvec$ for interruptive event of type $j$, and $p_j(\zvec; \betavec)$ can be modeled as
$p_j(\zvec; \betavec_j)=\exp(\zvec'\betavec_j)/(1+\exp(\zvec'\betavec_j)).$
From the model, AI algorithms with good internal reliability properties have small $p_j(\zvec;\betavec_j)$, and thus are less influenced by the interruptive events. However, more sophisticated statistical models and test data are needed for AI reliability research.

\subsubsection{Out-of-Distribution Detection}

\begin{figure}
\centering
\includegraphics[width = 0.8\textwidth]{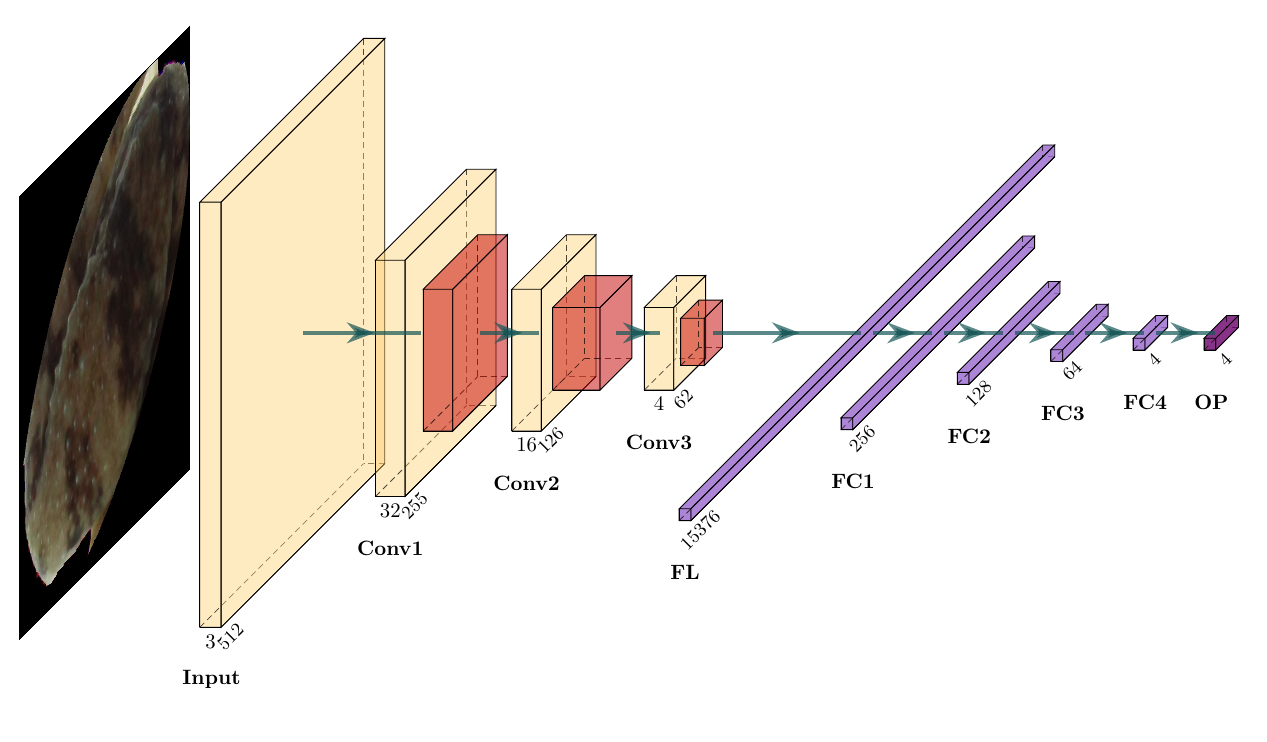}
\caption{Illustration of the CNN network architecture for frog leg image classification. \emph{Figure reproduced with permission from Springer}. }\label{fig:frogleg}
\end{figure}

Out-of-distribution (OOD) detection is one well known research problem that is related to the robustness of AI systems. Applied statistics can contribute to the detection of OOD and thus improve the robustness of AI models. For example, \shortciteN{Xuetal2023-JABES} proposed an algorithm of new class detection in neural networks utilizing the outputs from intermediate layers. In particular, the paper developed an algorithm in classifying the images of frog legs in differentiating species of \emph{Limnonectes kuhlii} complex (e.g., \shortciteNP{mcleod2010least}). The convolutional neural network (CNN) used in \shortciteN{Xuetal2023-JABES} is illustrated  in Figure~\ref{fig:frogleg}.

In the proposed OOD algorithm, linear discriminate analysis and quadratic discriminant analysis on $\zvec_i$ are first performed. Here, $\zvec_i$ denotes intermediate outputs from fully connected layers. A distance of $\zvec_i$ with its nearest class is then measured based on Mahalanobis distance. To achieve this, the paper assumed that the conditional distribution of $\zvec_i$ given class $j$ is a multivariate normal distribution with mean $\muvec_j$ and covariance matrix $\Sigma_j$, $j = 1,2, \dots, m$, where $m$ is the number of frog classes. The Mahalanobis distance-based confidence score $w_i$ is then calculated for $\zvec_i$. The paper then performed logistic regression with covariate $w_i$'s to detect whether an image should be in the OOD set. The paper demonstrated the success of the proposed new class detection using the MNIST dataset.

\subsubsection{Reliability Analysis of Autonomous Vehicles}

\begin{figure}
	\centering
	\begin{tabular}{cc}
		\includegraphics[width=.48\textwidth]{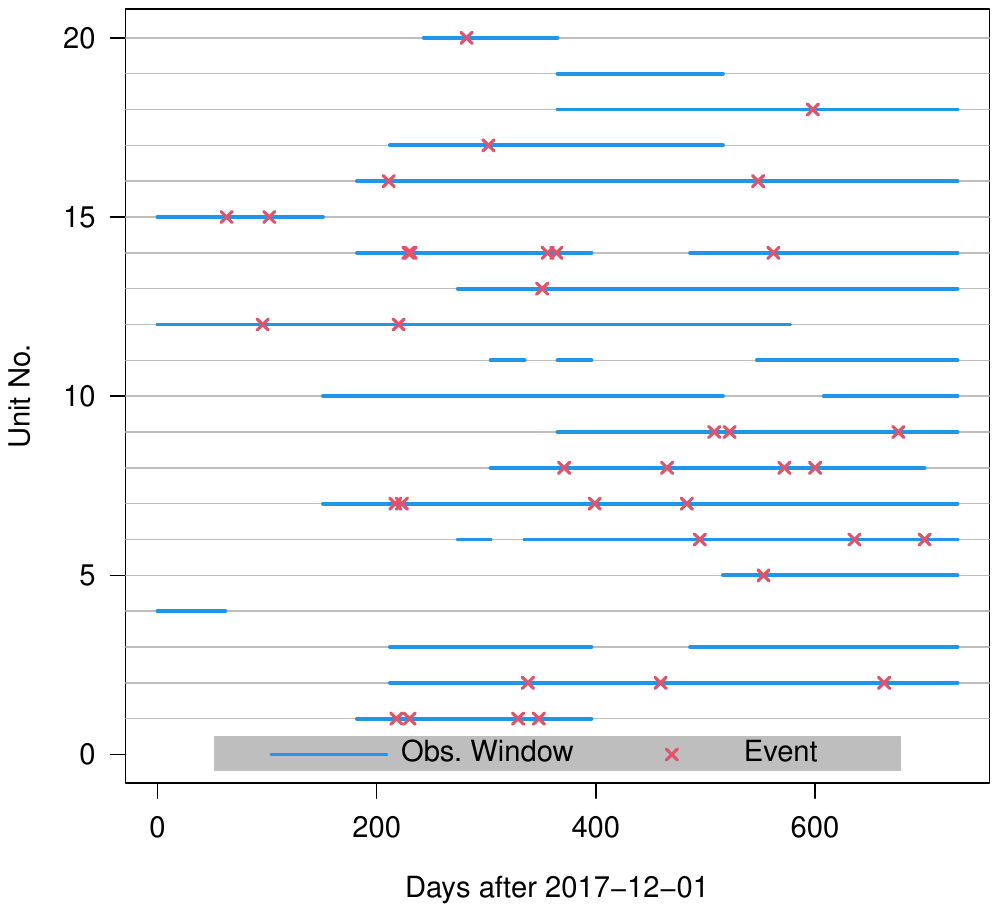}&
        \includegraphics[width=.48\textwidth]{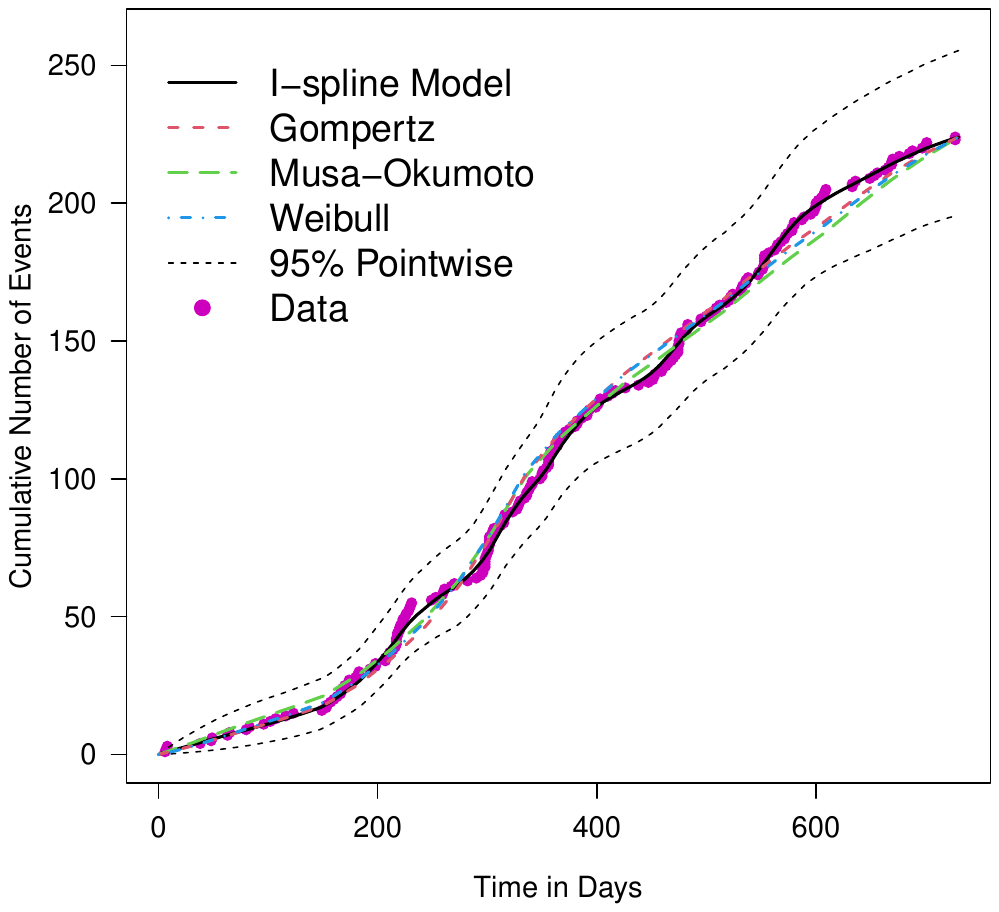}\\
		(a) Disengagement Events & (b) Observed vs. Expected\\
	\end{tabular}
	\caption{Example of recurrent events data from Waymo~(a), and plot of goodness of fit for various models~(b). \emph{Figure reproduced with permission from Oxford University Press}.}\label{fig:sample}
\end{figure}

Alternatively from the model in \eqref{eq:lambda1}, \shortciteN{MinHongKingMeeker2020} modeled the failure events of autonomous vehicles (AVs) using a nonhomogeneous Poisson process model. The paper analyzed the disengagement events data of AVs from four manufacturers (i.e., Waymo, Cruise, PonyAI, and Zoox).  Figure~\ref{fig:sample}(a) shows the disengagement events of $20$ vehicles.

In \shortciteN{MinHongKingMeeker2020}, the intensity function of failure events from vehicle $i$ is modeled as
\begin{equation*}
	\lambda_i[t; \xvec_i(t), \thetavec] = \lambda_0 (t; \thetavec) x_i(t), \quad i = 1,2,\dots,n,
\end{equation*}
where $\lambda_0(t;\thetavec)$ is the baseline intensity function with parameters $\thetavec$, $x_i(t)$ is the daily driven mileage of vehicle $i$, and $n$ is the total number of vehicles. I-spline bases are then used to model the cumulative baseline intensity function $\Lambda_0(t; \thetavec)$, which is
\begin{equation*}
\Lambda_0(t;\thetavec) = \int_{0}^t \lambda_0(s;\thetavec)ds=\sum_{l=1}^{n_{s}}\beta_l\gamma_{l}(t), \quad \beta_l\geq 0,\, l=1,\ldots, n_s,
\end{equation*}
where $\gamma_l(t)$ are the spline bases, $\beta_l$'s are the corresponding coefficients, and $\thetavec = (\beta_1, \dots, \beta_n)^\prime$. Confidence intervals and confidence bands of the cumulative intensity function of the spline model are developed based on random weighted bootstrap. Besides the flexible spline model, traditional parametric software reliability models are also used as complimentary tools. The parametric models included in the analysis are the Gompertz model, Musa-Okumoto model, and Weibull model. Figure~\ref{fig:sample}(b) shows the analysis results from spline and parametric models based on data from Waymo. The proposed statistical modeling framework can also be extended to analyze reliability of AI systems other than AVs.

\subsubsection{AI Test Planning}
Regarding test planning for AI systems, \shortciteN{hong2023statistical} discussed using the design of experiments and accelerated tests for evaluating AI systems. \shortciteN{hong2023statistical} mentioned acceleration methods such as increasing user rate, adding noise to training data, and error injection as effective strategies for accelerated testing of AI algorithms. For assurance testing, \shortciteN{Zheng2023-testplan} proposed a comprehensive planning method based on AV disengagement data. This method considers multiple criteria, including consumer's risk, producer's risk, testing time, and acceptance probability (the likelihood of passing the test), with consumer's risk being prioritized as the most important criterion. Disengagement events for AVs are assumed to follow a Poisson process, and a Bayesian framework is employed to calculate posterior risks based on the statistical model and data. A modified Pareto front approach is used to identify optimal test plans, making the framework useful for AI test planning with multiple objectives.

\section{AI for Applied Statistics}\label{sec:AI.for.AS}

Just as statistics has the potential to support and assist AI, so AI has the potential to support and assist applied statistics. In this section, we discuss ways in which AI can be valuable for statistics.

\subsection{AI Models for Applied Statistics}

AI models and technology are widely incorporated into modern applied statistics to address and overcome the limitations of traditional statistical methods. AI models excel at making classifications and predictions, aligning well with key goals of traditional statistical analysis. Here, we present an example from a recent study.

Electroluminescence (EL) images are widely used to study the reliability of PV modules in the field. EL images can be used to classify the PV modules into categories such as functional, mildly defective, moderately defective, and severely defective categories. \shortciteN{song2024comprehensivecasestudyperformance} applied both traditional machine learning techniques, including logistic regression, support vector machines (SVM), random forests (RF), and modern deep learning models like VGG19 and ResNet50 (modified to fit the dimensions of EL image data) to detect the severity of defects in solar cells through image classification. The data used in the paper includes 2624 EL solar cell images from both monocrystalline and polycrystalline PV modules labeled by experts (e.g., \shortciteNP{Deitsch2019}). Each image is 300 $\times$ 300 pixels in greyscale format. Images are normalized and augmented by randomly flipping, rotating, and adding Gaussian noise to prepare for the machine learning/deep learning models.

\begin{figure}
\begin{center}
\begin{tabular}{cc}
\includegraphics[width=0.45\linewidth]{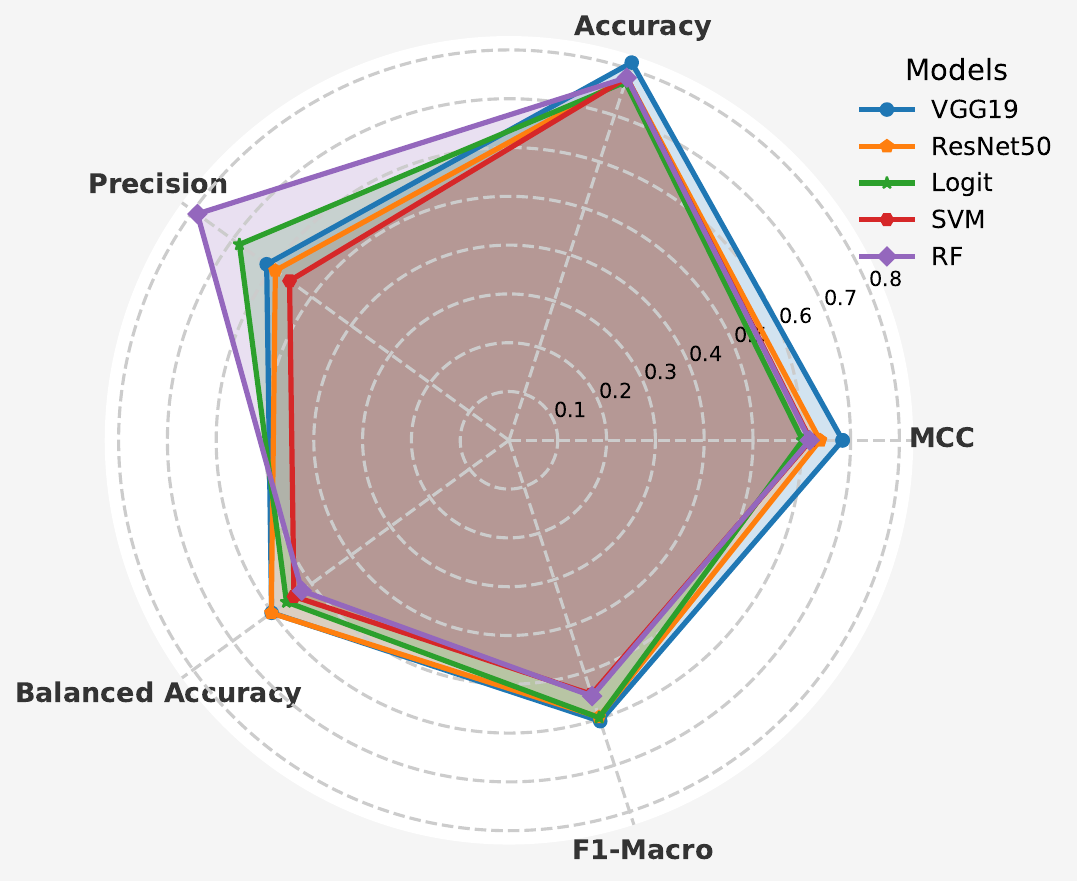}&
\includegraphics[width=0.4\linewidth]{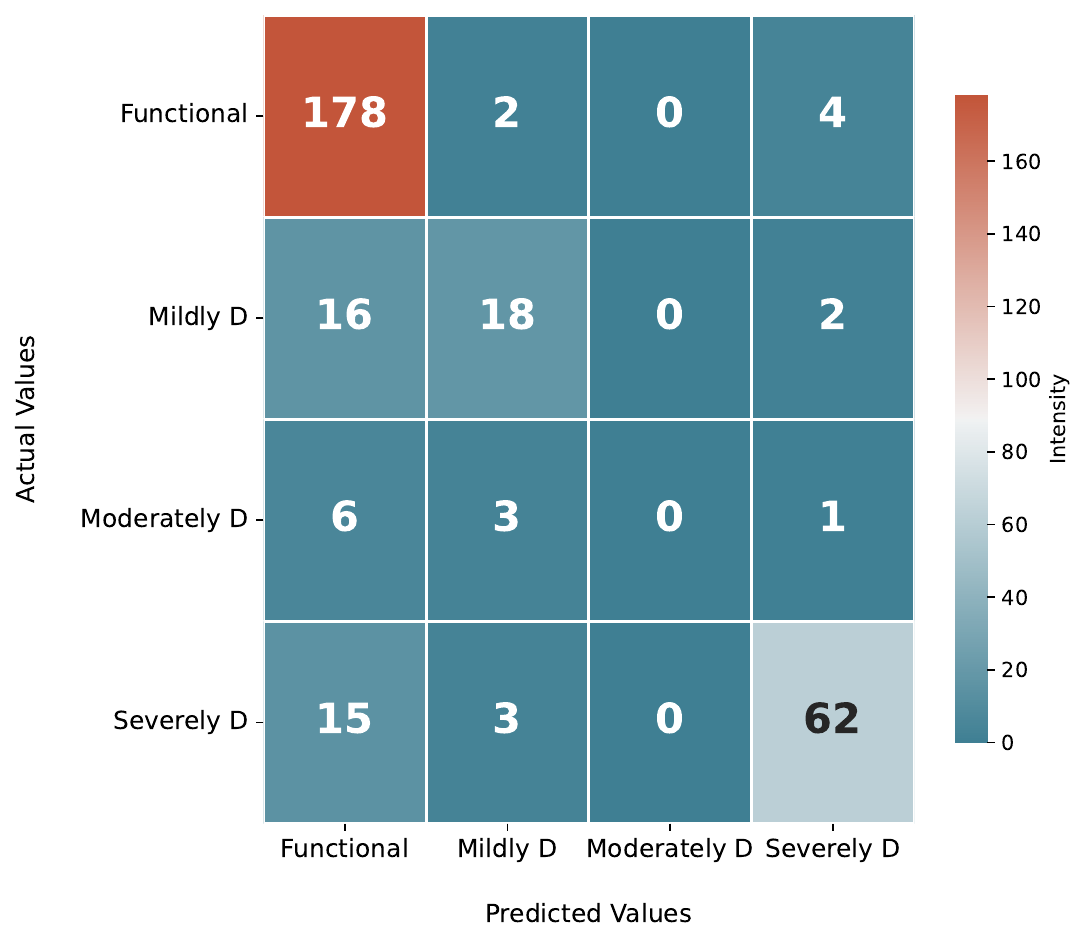}\\
(a) & (b)
\end{tabular}
\caption{Visualization of the median metrics for various machine learning models and confusion matrix in classifying EL images from PV modules.} \label{fig:mono.el.median.poly_cm}
\end{center}
\end{figure}

Figure~\ref{fig:mono.el.median.poly_cm}(a) gives the median values across multiple evaluation metrics such as balanced accuracy, precision and so on for machine learning/deep learning models among 50 replicates. The plot illustrates that a model that achieves high overall accuracy may not necessarily perform well in terms of balanced accuracy especially in scenarios under imbalanced datasets. In such cases, accurately predicting the minority class becomes difficult when equal weight is given to each class. Figure~\ref{fig:mono.el.median.poly_cm}(b) presents the confusion matrix for the prediction results of the ResNet-50 model for polycrystalline PV modules: while the model generally performs well, it does not effectively predict minority classes, such as mildly defective and moderately defective categories. This happens because models trained on such imbalanced datasets tend to prioritize accuracy for the majority class while neglecting the minority class (\shortciteNP{shyalika2023comprehensivesurveyrareevent}).

In another example of AI applications in applied statistics fields, \shortciteN{colosimo2021artificial} explored the use of AI in quality technology, addressing data challenges in quality monitoring and how AI models can help overcome them. The paper examines connections between design of experiments and AI topics, such as active learning and A/B testing. It also discusses the role of AI in predictive modeling of quality data and reliability analysis. Recently, \shortciteN{Megahedetal2024} introduced ChatSQC, a chatbot system that combines large language models (LLMs) with statistical quality control, further emphasizing the value of integrating AI into applied statistics to leverage the strengths of both fields.

\subsection{AI-Assisted Statistical Analysis}
In recent years, the rise of LLMs has led to substantial progress in AI-powered coding tools like, Copilot and ChatGPT. It also has expanded the possibilities for AI-assisted statistical analysis, which can harness the power of AIs to automate and enhance data analysis in both efficiency and accuracy, allowing people to focus more on strategy and decision making instead of tedious data manipulation tasks. \shortciteN{hong2024datainterpreterllmagent} introduced the data interpreter, an LLM-based agent designed to address challenges related to real-world data adaptability, tool integration, and data generation. It has been evaluated through various data science and real analysis tasks, demonstrating superior performance compared to open-source baselines, such as GPT-4-Turbo.

Recent studies have explored the effectiveness of AI-assisted data analysis across various domains, highlighting the potential for growth and innovation in this emerging area. \citeN{bdcc7020062} investigated the role of ChatGPT in data science, providing examples that showcase its statistical capabilities. The study found that while ChatGPT performs well in training models from scratch and fine-tuning, it struggles with tasks in untrained domains, and its results can be challenging to interpret. Other studies examine ChatGPT's data analysis abilities within specific datasets or domains. For example, \citeN{AIMedicine2024} assessed ChatGPT's accuracy in answering common cardiac-related medical questions, finding that the latest version shows strong potential for handling straightforward, simple medical inquiries. Based on 50 cardiovascular trivia questions and 20 clinical case vignettes, the study noted that further research is necessary to fully assess ChatGPT's accuracy and reliability.

In addition to research studies, some real-world AI-assisted data analysis products have already been released. Although still in its developmental stages, there are platforms that integrate AI and machine learning into data science and business analytics. However, it remains unclear whether the accuracy and reliability of those analysis results can be assured. Nonetheless, we expect more tools like this to be developed in the future.

We also want to highlight the potential of AI-driven data augmentation, which represents a promising avenue for future advancements in applied statistics. As AI continues to evolve into an extensive knowledge base, it has the capacity to provide supplementary information that can significantly enhance data augmentation techniques in statistical analysis.

\subsection{Computing and Software}

Traditional statistical software such as SAS, JMP, Stata, Python, and R are now incorporating AI to improve their capabilities. These enhancements include using LLMs to help with code writing, debugging, programming language conversion with AI assistance such as converting SAS into Python, which makes these tools more efficient and powerful. Many studies have investigated the accuracy of AI-generated code (\shortciteNP{COTRONEO2024112113}; \shortciteNP{Clark2024AIConsistency}; \shortciteNP{poldrack2023aiassistedcodingexperimentsgpt4}). \citeN{codegenerate2024} evaluated the Java, Python and C++ code generated by multiple LLMs such as BingAI Chat (GPT-4.0), ChatGPT (GPT-3.5), Code Llama (Llama 2) and others in terms of time and space complexity, runtime and memory usage, correctness, efficiency and maintainability. Their study showed that AI-generated code works well with easy problems but suffers from inefficiencies, bugs, or incorrect logic when dealing with more difficult tasks. Additionally, several studies have highlighted security and ethics concerns with AI-generated code, particularly regarding improper input validation, and security vulnerabilities (\shortciteNP{cotroneo2024devaictoolsecurityassessment}; \shortciteNP{wang2024aigeneratedcodereallysafe}; \shortciteNP{insecureai2023Perry}; \shortciteNP{XuSheng2024AIdetect}; \shortciteNP{AIethics2022BNS}). \shortciteN{kaniewski2024vulnerabilityhandlingaigeneratedcode} examined the current landscape of LLM-based methods for code vulnerability handling and pointed out some challenges of vulnerability detection, localization, and repair. However, AI-generated code also carries the potential to enhance statistics and data science. \shortciteN{GPTForeducation2023Amanda} noted that AI tools such as ChatGPT can help generate course materials focusing on statistical programming, and can also write code for the same tasks across multiple programming languages, including R, Python, and SAS. This capability promotes code generalization and compatibility, while also enhancing collaboration in both academia and industry.

Human communication with statistical software has evolved through three distinct generations, each reflecting advancements in technology and user experience.

\begin{inparaitem}
\item The first generation was characterized by command-line interfaces or compiled languages, requiring users to input precise commands and scripts to perform statistical analyses, often demanding a high level of technical proficiency.

\item The second generation introduced graphical user interfaces (GUIs), allowing users to interact with software through menus and clicks, which made statistical tools more accessible to non-experts and enabled a more intuitive approach to data analysis.

\item The third generation represents a significant leap forward, leveraging LLM to enable users to communicate with statistical software using everyday language. Although this stage is still under development, this evolution not only democratizes access to complex statistical methods but also enhances usability, as users can now engage with the software in a more conversational and fluid manner.
\end{inparaitem}

\section{Outlook and Concluding Remarks}\label{sec:concluding.remarks}

In this section, we discuss the evolving role of applied statisticians and offer some concluding reflections.

\subsection{Automation in Statistical Analysis}

Current AI developments, such as ChatGPT, can assist in statistical analysis, though they have not yet reached the level of full automation. However, the trend is moving towards developing automation in statistical analysis, creating a kind of statistics robot (stat-bot). In the future, an AI-powered framework for automatic statistical analysis will likely become available, which involves steps 3 to 7 in Figure~\ref{fig:apllied.stat.flowchart}.

In statistical practice, significant training is required for individuals to perform statistical analysis. This training includes learning statistical concepts, methods, software implementations, content organization, and proper deduction. Therefore, developing an automatic statistical analysis framework is desirable. The user of such a system would need to provide a problem description, a data description, and the dataset. The stat-bot would then undertake a series of automated tasks, such as understanding the problem's objective, executing a statistical program in software like SAS or R, and ultimately presenting the analysis results. In this way, the stat-bot can understand the problem, conduct the analysis, and present the results.

Automating statistical analysis with AI is challenging because it involves a complex system, including problem description, data description, dataset selection, modeling, programming, analysis, and reporting. Achieving automation in statistical analysis requires developing an integrated framework of AI-powered automation for extracting key information, learning the analysis, forming executable code, summarizing results, and generating a written report.

\subsection{The Changing Roles of Applied Statisticians}
With all these exciting developments of AI for statistical analysis, the role of statisticians is likely to change in the future. Though AI will automate certain aspects of data analysis and predictive modeling, statisticians will continue to be indispensable for their analytical skills, domain expertise, and ability to ensure the integrity and interpretability of data-driven insights.

The future of statistical practice will likely involve a symbiotic relationship between AI and human statisticians, leveraging the strengths of both to advance research, innovation, and decision-making across various fields. AI is unlikely to completely replace statisticians, but it will certainly change the nature of their roles and the tools they use.

Statisticians have been transitioning from traditional roles to becoming data scientists with an applied statistics background. Data scientists with this background are well-equipped to handle the challenges of data science, from data cleaning and analysis to modeling and interpretation. Their expertise in statistics provides a strong foundation for building robust data-driven solutions. However, statisticians often face challenges with domain knowledge and programming skills. The development of AI tools for statisticians is definitely helpful for these transitions.

The development of AI also provides tremendous opportunities for statistical research, similar to those we have seen with other emerging technologies, such as bioinformatics. In AI-related research, a statistician can be a leader rather than just a collaborator. For example, since many AI tools are essentially algorithms, their robustness and safety can be studied using computing and simulation. Thus, statisticians can be more involved in the design of experiments, data collection, and data analysis.

\subsection{Concluding Remarks}

AI is transforming nearly every industry, including applied statistics, and bringing both opportunities and challenges. While traditional roles may be phased out, new job opportunities are being created in their place. As the saying goes, ``It was the best of times, it was the worst of times." Should we feel more concerned than excited about AI's growing influence? As outlined in this paper, AI promises to bring tremendous opportunities to the field of applied statistics. It's crucial to remain positive and open-minded about these new developments.

While AI continues to advance and automate an increasing number of tasks, certain human skills remain uniquely difficult to replicate. These skills often encompass human intuition, ethical judgment, complex emotional intelligence, and creativity, areas where AI currently falls short. For future statisticians, focusing on these irreplaceable skills is crucial, as they are the foundation of innovative problem-solving and responsible decision-making in the face of uncertainty. Training programs for statisticians should emphasize the development of these competencies, enabling them to complement AI technologies effectively and navigate the nuanced challenges that automated systems alone cannot address. This human-centered approach will not only set future statisticians apart but also ensure that they remain at the forefront of their field in an increasingly AI-driven world.

Finally, we want to acknowledge the limitations of this work. Given its broad scope, our literature review is unlikely to be exhaustive. The field of applied statistics is vast and its boundaries are not rigorously defined, constantly evolving over time. Rather than striving for completeness, we focused on sharing our perspectives and experiences. Our illustrative examples are centered on engineering statistics, an area with which we are most familiar. However, we recognize that the applications of applied statistics extend far beyond this domain.

\section*{Acknowledgments}

The work by Deng and Hong was supported in part by the COS Dean's Discovery Fund at Virginia Tech (Award: 452021). The work by Hong was supported in part by the Data Science Faculty Fellowship (Award: 452118) at Virginia Tech.


\end{document}